\documentclass[sigconf,nonacm,screen]{acmart}
\AtBeginDocument{%
  }

\setcopyright{acmlicensed}
\copyrightyear{2018}
\acmYear{2018}
\acmDOI{XXXXXXX.XXXXXXX}
\acmConference[Conference acronym 'XX]{Make sure to enter the correct
  conference title from your rights confirmation email}{June 03--05,
  2018}{Woodstock, NY}
\acmISBN{978-1-4503-XXXX-X/2018/06}

\usepackage{url}
\usepackage{lipsum}
\usepackage{
  float,
  epsfig,
  wrapfig,
  graphicx,
  subcaption
}
\usepackage{colortbl}
\usepackage{enumitem}
\usepackage{graphics}
\usepackage{xparse}
\usepackage{bbding}
\usepackage{pifont}
\usepackage{xspace}
\usepackage{wasysym}
\usepackage{multirow}
\usepackage[dvipsnames]{xcolor}
\usepackage{diagbox}
\usepackage{booktabs}
\usepackage{makecell}
\usepackage{listings}
\usepackage{hyperref}
 
\usepackage{amssymb}
\usepackage{amsmath}
\usepackage{algorithm}
\usepackage[noend]{algcompatible}

\usepackage{soul}

\def\eg{\emph{e.g.}\xspace}
\def\ie{\emph{i.e.}\xspace}

\newcommand{\tool}{\textsc{Heimdallr}}

\newcommand{\as}{\textit{Model-agnostic auditing}}

\begin{document}

\title{An Effective and Cost-Efficient Agentic Framework for Ethereum Smart Contract Auditing}

\author{Xiaohui Hu}
\authornote{Both authors contributed equally to this research.}
\affiliation{%
  \institution{Huazhong University of Science and Technology}
  \country{China}
}

\author{Wun Yu Chan}
\authornotemark[1] 
\affiliation{%
  \institution{Amber Group}
  \country{Singapore}
}

\author{Yuejie Shi}
\affiliation{%
  \institution{Amber Group}
  \country{Singapore}
}

\author{Qumeng Sun}
\affiliation{%
  \institution{University of Göttingen}
  \country{Germany}
}

\author{Wei-Cheng Wang}
\affiliation{%
  \institution{National Yang Ming Chiao Tung University}
  \country{Taiwan, China}
}

\author{Chiachih Wu}
\affiliation{%
  \institution{Amber Group}
  \country{Singapore}
}

\author{Haoyu Wang}
\affiliation{%
  \institution{Huazhong University of Science and Technology}
  \country{China}
}

\author{Ningyu He}
\affiliation{%
  \institution{The Hong Kong Polytechnic University}
  \country{Hong Kong SAR, China}
}

\begin{abstract}
Smart contract security is paramount, but identifying intricate business logic vulnerabilities remains a persistent challenge because existing solutions consistently fall short: manual auditing is unscalable, static analysis tools are plagued by false positives, and fuzzers struggle to navigate deep logic states within complex systems. Even emerging AI-based methods suffer from hallucinations, context constraints, and a heavy reliance on expensive, proprietary Large Language Models.
In this paper, we introduce {\tool}, an automated auditing agent designed to overcome these hurdles through four core innovations. By reorganizing code at the function level, {\tool} minimizes context overhead while preserving essential business logic. It then employs heuristic reasoning to detect complex vulnerabilities and automatically chain functional exploits. Finally, a cascaded verification layer validates these findings to eliminate false positives. Notably, this approach achieves high performance on lightweight, open-source models like GPT-oss-120B without relying on proprietary systems.
Our evaluations demonstrate exceptional performance, as {\tool} successfully reconstructed 17 out of 20 real-world attacks post June 2025, resulting in total losses of \$384M, and uncovered 4 confirmed zero-day vulnerabilities that safeguarded \$400M in TVL. Compared to SOTA baselines including both official industrial tools and academic tools, {\tool} at most reduces analysis time by 97.59\% and financial costs by 98.77\% while boosting detection precision by over 93.66\%. Notably, when applied to auditing contests, {\tool} can achieve a 92.45\% detection rate at a negligible cost of \$2.31 per 10K LOC. We provide production-ready auditing services and release valuable benchmarks for future work.
\end{abstract}



\received{20 February 2007}
\received[revised]{12 March 2009}
\received[accepted]{5 June 2009}

 \maketitle

\section{Introduction}
The exponential proliferation of Decentralized Finance (DeFi) has revolutionized the financial landscape, yet this rapid expansion has been accompanied by a surge in severe security incidents. According to the report~\cite{certik2025}, there were over 700 security incidents, leading to total losses of approximately \$3.35 billion, a 66.64\% increase from the previous year.
Smart contracts, the immutable code governing these protocols, manage over \$123.6 billion in assets~\cite{defi-statistic}, making them prime targets for malicious exploitation. 

From the security side of the implementation of DeFi protocols, most of existing work focus on conducting static analysis or dynamic testing on deployed on-chain ones.
For example, Zhang's work~\cite{bosi2025following} identifies price manipulation attack via analysis on bytecode, and Sun's work~\cite{sun2024allyoutokens} conducts strict checks in the address verification process.
However, such methods have an inherent limitation: the deployment process, as well as the identified vulnerabilities, cannot be reversed.

\begin{figure}
    \centering
    \includegraphics[width=0.8\linewidth]{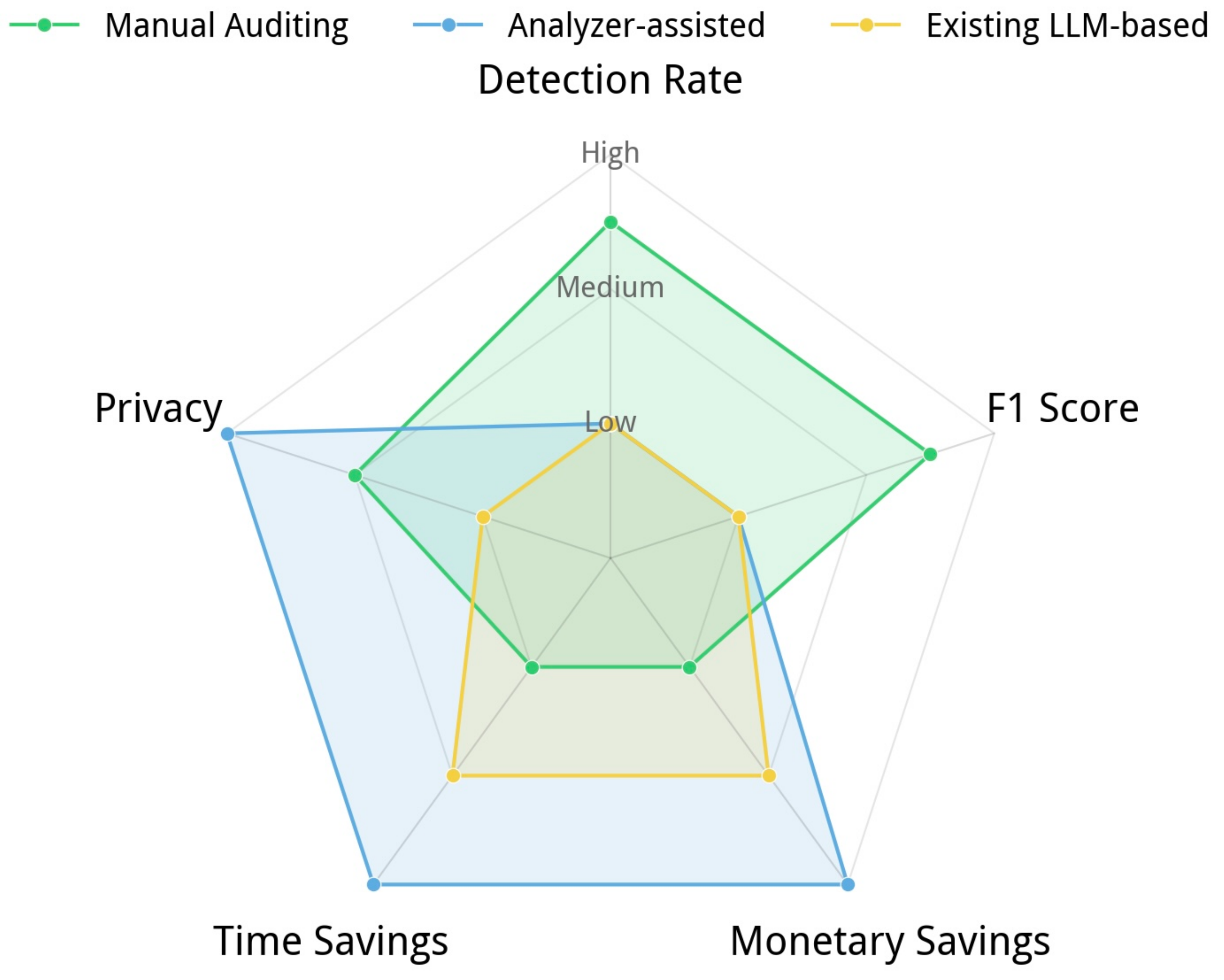}
    \caption{Qualitative evaluation of existing smart contract auditing strategies.}
    \label{fig:qualitative evaluation of three strategies}
\end{figure}

Auditing somehow alleviates the above limitation as it is conducted \textit{before} the deployment. Contract auditing could be divided into three mainstream categories, \textit{i.e.,} \textit{manual auditing}, \textit{analyzer-assisted auditing}, and \textit{LLM-based auditing}, Figure~\ref{fig:qualitative evaluation of three strategies} qualitatively characterizes them from different perspectives.
Specifically, manual auditing is the most mainstream way, which, however, is prohibitively expensive and unscalable. Reports suggest a medium DeFi protocol will require \$70,000–\$150,000+ for auditing~\cite{manualAuditing}.
Regarding analyzer-assisted auditing, existing SOTA academic analyzers, like Slither~\cite{slither} and Mythril~\cite{mythril}, and industrial self-developed closed-source analyzers are used to \textit{assist} manual auditing. However, analyzers often suffer from extremely limited scope and high false positives, cannot fully replace manual auditing at all~\cite{chaliasos2024smart, perez2021smart}.
Recently, Large Language Models (LLMs) introduce new possibilities due to their increasing context window and reasoning capabilities~\cite{ma2024combiningfinetuningllmbasedagents, sikder2025efficient}. Some academic work~\cite{wei2024llmaudit, sun2024gptscan, ma2024combiningfinetuningllmbasedagents} and industrial solutions~\cite{hound, hashlock} have adopted various LLMs as backends to conduct automatic contract auditing. However, they are plagued by hallucinations, leading to unmanageable false positive rates. Our work applied three existing LLM-based tools on 80 real-world protocols (\S\ref{sec:evaluation}), finding that they exhibit either negligible detection capabilities (F1 Score < 0.02) or prohibitive noise levels (FPR > 97\%), rendering them practically unusable for detecting DeFi exploits.

\textit{Is it possible to leverage the automation and reasoning capabilities of LLMs, and minimize false positives/negatives caused by hallucinations, to achieve scalable and highly reliable contract audit results?}
Motivated by this gap, this research focuses on two primary challenges, \textit{i.e.,} \textit{effective false positive reduction} and \textit{the optimization of cost-efficiency without compromising detection accuracy}. 

To resolve above challenges, in this work, we design {\tool} as a cohesive pipeline with three methodologies (\S\ref{sec:method}). Generally speaking, \textit{Contextual Profiling} optimizes context retention via graph clustering; \textit{Model-Agnostic Auditing} combines semantic and neuro-symbolic reasoning for comprehensive detection; and \textit{False Positive Filtration} rigorously filters hallucinations through adversarial feasibility checks, ensuring high-precision results at minimal cost.

\textbf{This work.}
We implement {\tool} as a modular, model-agnostic framework that integrates three core components: \textit{Profiler} for logic-preserving code batching, a ``Plan-Remind-Solve'' strategy adopted \textit{Auditor} for heuristic reasoning, and a \textit{Verifier} for adversarial filtration. Extensive evaluations demonstrate {\tool}’s superior effectiveness and efficiency. In reproducing high-value exploits post-June 2025, {\tool} successfully reconstructed 17 out of 20 attacks, totaling \$384 million in losses, and further identified four confirmed zero-day vulnerabilities in live protocols with \$400 million TVL. Compared to SOTA baselines including both popular industrial and academic tools, {\tool} at most reduces analysis time by 97.59\% and financial costs by 98.77\% while boosting detection precision by over 93.66\%. Notably, when applied on auditing contests, {\tool} can at most achieve a 92.45\% detection rate at a negligible cost of \$2.31 per 10K LOC, ranking 19th out of 548 participating human auditors.

\textbf{Our contributions.} We summarize our key contributions of this work as follows:
\begin{itemize}[leftmargin=*]
    \item \textbf{Novel Auditing Methodology.} We propose {\tool}, a framework that combines graph-theoretic profiling with neuro-symbolic reasoning to resolve the scalability-precision trade-off. Our ``Plan-Remind-Solve'' workflow detects complex logic vulnerabilities while the adversarial verification module rigorously suppresses false positives, ensuring high-fidelity results.
    
    \item \textbf{Production-Ready Auditing Service.} To balance utility and safety, we release {\tool} as a standardized API service for seamless integration into CI/CD pipelines. Additionally, we support privacy-preserving local deployment using open-source models, enabling secure auditing of proprietary codebases in air-gapped environments.
    
    \item \textbf{Comprehensive Empirical Evaluation.} We benchmark {\tool} against four state-of-the-art baselines, demonstrating a superior F1 score of 0.62 on real-world datasets, orders of magnitude higher than existing academic prototypes. Furthermore, we provide the first rigorous analysis of the precision-recall trade-off in LLM-based auditing, validating the critical necessity of decoupled verification.
    
    \item \textbf{High-Fidelity Vulnerability Benchmarks.} We construct and release three distinct datasets covering high-value zero-day exploits, standardized benchmarks, and crowdsourced contest findings. These resources address the scarcity of diverse ground-truth data in smart contract security research, facilitating future comparative studies.

\end{itemize}

\textbf{Tool Usage Request.}
To prevent potential misuse by malicious actors, we have decided not to open-source the audit engine. However, researchers and practitioners interested in utilizing our tool can request access to our API-based audit service by contacting Xiaohui Hu (xiaohui\_hu@hust.edu.cn), Chiachih Wu (chiachih.wu@ambergroup.io), or Ningyu He (ningyu.he@polyu.edu.hk).
\section{Background}

\subsection{Smart Contract Vulnerabilities in DeFi}
Smart contracts are self-executing programs underpinning Decentralized Finance (DeFi) on platforms like Ethereum~\cite{Sharma2023}. DeFi protocols provide diverse financial services without intermediaries, including decentralized exchanges (DEXs)~\cite{dex}, lending~\cite{lending}, stablecoins~\cite{klages2020stablecoins}, portfolio management~\cite{platanakis2019portfolio}, and derivatives~\cite{jensen2021introduction-derivative}. We exclude privacy-focused mixers~\cite{werner2022sok, yufeng2025mixer} from our scope due to their association with illicit activities.

In this work, we classify smart contract vulnerabilities into two primary categories:
\textit{Non-business logic vulnerabilities} are generic security flaws inherent to the code structure, independent of specific protocol functionality. Examples include reentrancy, integer overflows/underflows, access control weaknesses, and unprotected delegate calls~\cite{zhou2023sokdecentralizedfinancedefi}. These vulnerabilities typically stem from language-specific pitfalls (\eg, Solidity quirks) and follow recognizable anti-patterns, making them detectable via pattern-matching static analysis tools like Slither~\cite{slither}.
\textit{Business logic vulnerabilities} arise from flaws in protocol-specific economic mechanisms or operational workflows. Examples include incorrect incentive structures enabling economic exploits, oracle manipulation, inadequate collateralization checks, and state transition errors that violate protocol invariants~\cite{zhou2023sokdecentralizedfinancedefi}. Unlike generic bugs, these vulnerabilities are highly context-dependent. Detecting them requires a deep semantic understanding of the protocol's intended financial model and cross-contract interactions, rendering traditional pattern-matching approaches ineffective.

\subsection{LLM-Driven Autonomous Agents}
Large Language Models (LLMs) are AI systems trained on massive corpora to encode linguistic structure and world knowledge via next-token prediction~\cite{gpt2, gpt3}. Beyond pretraining, advanced post-training techniques like instruction tuning and reinforcement learning~\cite{rlhf} have significantly enhanced their reasoning capabilities, enabling complex, step-by-step problem solving~\cite{guo2025deepseek}.
Building on this foundation, \textbf{LLM Agents} extend pure text generation into goal-directed systems capable of autonomous multi-step task execution~\cite{qiao2024autoact}. An agent architecture wraps the LLM with modules for planning~\cite{yao2022react}, memory management~\cite{chhikara2025mem0}, and external tool access~\cite{schick2023toolformer}. Unlike a standard LLM that merely outputs text, an agent utilizes the model as a reasoning core to formulate plans, invoke APIs, and verify outcomes~\cite{qin2024toollearning}. For example, rather than hallucinating weather data, an agent can autonomously query a weather API to provide factual results.
To facilitate seamless tool integration, the \textbf{Model Context Protocol (MCP)} has emerged as an open standard for secure context exchange~\cite{anthropic2024b}. MCP provides a vendor-neutral interface that allows agents to discover and utilize external resources—such as file systems, databases, or web search APIs—without requiring bespoke connectors. Operating on a client-server model via JSON-RPC 2.0, MCP enables an AI host to dynamically connect to MCP servers, discover capabilities, and execute standardized tool calls over transports like HTTP/SSE. This standardization is critical for building modular, extensible auditing agents that can safely interact with diverse external knowledge bases and analysis tools.
\section{Motivation \& Challenges}
In this section, we first summarize the limitations of existing tools with a typical DeFi development scenario. 
Then, we summarize and present two major challenges to be solved.

\subsection{Current Limitations}
Consider a smart contract development team preparing to deploy a decentralized finance (DeFi) protocol on the Ethereum mainnet. 
As the project approaches production deployment, the team confronts a critical security imperative, \textit{i.e.,} \textit{\ul{ensuring that their contracts are free from vulnerabilities that could result in malicious exploitations}}.
Three strategies naturally come to mind, \textit{i.e.,} \textit{manual auditing}, \textit{analyzer-assisted auditing}, and \textit{LLM-based auditing}.

\textit{\textbf{Manual Auditing.}}
Traditional manual auditing presents significant barriers, where high costs and quality uncertainties pose the primary concerns. 
Engaging professional auditing firms typically costs approximately \$2,500 for a modest codebase of around 1,000 lines of code~\cite{manualAuditing}, while turnaround times range from one to two weeks, depending on auditor availability and project complexity.
Quality variation across firms compounds the cost challenge. Top-tier auditors deliver comprehensive reports. Their analyses include a deep examination of business logic vulnerabilities. In contrast, less experienced auditors often focus on surface-level, pattern-based issues. They frequently miss context-dependent vulnerabilities that require deeper protocol understanding.
Early-stage projects face particular difficulties with this approach. Limited budgets constrain their options. Tight launch schedules add pressure. Together, these costs and quality uncertainties create substantial friction in the development process.

\textit{\textbf{Analyzer-assisted Auditing.}}
Analyzer-assisted auditing employs automated security tools to support smart contract vulnerability detection. Recent work suggests these tools exhibit significant limitations that hinder their effectiveness in modern DeFi environments. 
For example, 38.1\% of auditors reported that existing tools were unhelpful~\cite{chaliasos2024smart}.
The primary flaw lies in their insufficient ability to capture complex business logic vulnerabilities~\cite{perez2021smart}. Logic errors depend entirely on the specific context and intent of the protocol. These vulnerabilities rarely manifest as generic code patterns. Consequently, static analyzers frequently miss the intricate state manipulations that cause major financial exploits.
Objectively, current tools exhibit suboptimal performance. An analysis of 127 real-world attacks, totaling \$2.3 billion in losses, reveals that existing tools detect vulnerabilities in only 25\% of cases~\cite{chaliasos2024smart}. This coverage accounts for merely 12\% of total damages (\$271 million).

\textit{\textbf{LLM-based Auditing.}}
LLM-based auditing also faces critical limitations that hinder practical adoption. Challenges arise from both \textit{direct LLM usage} and \textit{specialized agent frameworks}.
Specifically, direct LLM application (\eg, uploading contracts to online LLMs) presents three fundamental issues:
\begin{itemize}[leftmargin=*]
\item \textit{Incompleteness} occurs because models analyze only code visible within their context window. This restriction causes them to miss external inter-contract dependencies and multi-file call chains.
\item \textit{Usability friction} emerges from the need for manual file uploads across sessions. Users lack systematic coverage tracking throughout the analysis process.
\item \textit{Output instability} results from LLM non-determinism. The same code analyzed multiple times produces inconsistent findings.
\end{itemize}
Existing auditing agents, while promising automation, introduce their own distinct challenges:
\begin{itemize}[leftmargin=*]
\item \textit{Privacy concerns} arise with closed-source tools~\cite{hashlock}, which require transmitting sensitive code to third-party servers. This creates unacceptable intellectual property risks for many projects.
\item \textit{Practical unavailability} is derived since we find many existing tools are not publicly available for direct use, while some work~\cite{ma2024combiningfinetuningllmbasedagents, sikder2025efficient} often require complex self-deployment of fine-tuned large models. These factors create significant operational hurdles for average auditors.
\item \textit{False positive overload} plagues functional tools, with spurious warning rates exceeding 90\%~\cite{chaliasos2024smart}. Manual triage of these warnings becomes more burdensome than traditional auditing.
\item \textit{High operational costs} characterize cloud-based agents. A typical 10K-line project consumes \$15–\$50 in API expenses, with costs scaling linearly for larger codebases.
\end{itemize}
These scenarios illuminate the limitations of traditional non-LLM-based methods and the inability of LLM-based methods to achieve a balance among cost-efficiency, data privacy preservation, detection accuracy, and practical usability during smart contract auditing.

\subsection{Challenges}
\label{subsec: challenges}
To achieve greater scalability and versatility for the auditor, we still plan to adopt LLM-based solutions. Therefore, we need to address the following two challenges:

\label{challenge:1}
\noindent
\textbf{\textit{Challenge 1: Reducing False Positives.}}
LLM-based auditing tools frequently suffer from a high false positive rate, which primarily stems from three inherent limitations. 
First, LLMs are prone to hallucinations. They often report vulnerabilities that do not exist in the code. 
Second, these tools typically lack sufficient execution context. They cannot accurately track function call relationships across the codebase. 
Third, they often rely on unrealistic attack scenarios. For example, they may over-assume that an attacker possesses leaked private keys for privileged accounts. 
Consequently, \textit{\ul{many reported findings do not represent valid threats in real-world environments.}}
 
\label{challenge:2}
\noindent
\textbf{\textit{Challenge 2: Balancing Effectiveness and Cost-Efficiency.}}
Current LLM-based auditing tools face a severe trade-off between performance and cost. 
On the one hand, comprehensive analysis currently demands high-capability models like GPT-5. These models possess the reasoning power necessary to dissect complex smart contracts. However, this dependency incurs prohibitive expenses. A single audit project, in our practice, frequently costs between \$20 and \$50 due to extensive API token usage. This price point makes frequent auditing unsustainable for many developers.
On the other hand, attempts to reduce costs often compromise detection quality. Substituting smaller or open-source models results in drastic performance degradation. For instance, replacing GPT-5 with a model like GPT-oss-120B causes vulnerability detection recall to plummet by 60\% to 80\%. Small models struggle to maintain context and reason deeply. 
Consequently, auditors face a critical dilemma: \textit{\ul{unaffordable accuracy or unavoidable incompetence}}. The challenge lies in achieving excellent detection capabilities while utilizing cost-effective, resource-constrained models.

\section{Methodology}
\label{sec:method}
We design {\tool} as a cohesive pipeline that mimics the rigorous reasoning process of human experts through three logically coupled stages. 
First, \textit{Contextual Profiling} (\S\ref{sec:method:m1}) addresses the trade-off between scalability and context preservation. By constructing a function-level dependency graph and applying graph-theoretic clustering, this phase partitions the codebase into semantically coherent batches, ensuring that coupled business logic is grouped to maintain essential execution context. 
Next, these optimized batches feed into the \textit{Model-Agnostic Auditing} (\S\ref{subsec: methodology-diagnosis}) engine. Adopting a ``Plan-Remind-Solve'' agentic workflow, this phase integrates semantic reasoning with symbolic constraints and adversarial state analysis to generate a comprehensive set of vulnerability hypotheses ($\mathcal{V}_{raw}$). 
Finally, the process concludes with \textit{False Positive Filtration} (\S\ref{subsec: methodology-verify}). Functioning as an adversarial quality gate, this module validates the feasibility of each hypothesis against global safeguards and threat models, systematically pruning false positives to produce the final verified report ($\mathcal{V}_{final}$).

\subsection{Contextual Profiling}
\label{subsec: M1 methodology-profiler}
\label{sec:method:m1}
To enable scalable analysis without sacrificing cross-contract context, we develop a contextual profiling method to partition a given project into \textit{batches}. 
Formally, a project is a set of contracts, denoted as $\mathcal{C}$.
This method is designed to produce a set of batches, $\mathcal{B}=\{b_1,\dots,b_n\}$, where $b_i \subseteq \mathcal{C}$ such that $\bigcup b_i = \mathcal{C}$. 
Overlaps between batches are allowed to handle shared libraries or core utility contracts that are critical to multiple logical domains. 
The whole process can be divided into four steps.

\noindent
\textbf{Step 1: Static Graph Construction.}
Given the implementation of a project, we first use Slither~\cite{slither} to construct a static directed dependency graph: $\mathcal{G}=(\mathcal{F}_\mathcal{G}, \mathcal{E}_\mathcal{G}, \mathcal{W})$, where the node set $\mathcal{F}_\mathcal{G}$ comprises all \textit{functions} within the project.
The edge set $\mathcal{E}_\mathcal{G}$ captures two distinct types of interactions:
\begin{itemize}[leftmargin=*]
    \item \textbf{Control Flow ($E_{call}$)}: Edges $f_i \xrightarrow{call} f_j$ representing $f_j$ is invoked in $f_i$.
    \item \textbf{Data Flow ($E_{data}$)}: Edges $f_i \xrightarrow{\text{data}} f_j$ representing $f_j$ data-flow depends on $f_i$.
\end{itemize}

To quantify the coupling relationships, we assign weights $\mathcal{W}$ to edges.
Specifically, explicit control flow edges ($f_i \xrightarrow{call} f_j$) represent the direct dependency and are assigned a baseline weight $w_{call}=1.0$.
Regarding the data flow edges, here is an example: $f_i$ writes to a state variable $s_k$ and $f_j$ reads from $s_k$, we establish an edge $f_i \xrightarrow{\text{data}} f_j$. As this is an \textit{indirect} dependency between functions, we assign these edges a weight of $w_{state}=0.8$ (empirically tuned).

\noindent
\textbf{Step 2: Graph-Theoretic Community Detection.}
To identify logical boundaries for obtaining batches, we partition $\mathcal{G}$ into \textit{densely connected subgraphs}. 
We apply the Louvain Algorithm~\cite{louvain} to maximize the modularity of the weighted graph, yielding a set of communities $\mathcal{U}=\{U_1, \dots, U_m\}$, where each $U_j \subset \mathcal{V}_\mathcal{G}$.
Since functions in Solidity cannot be audited in isolation from their contract context, we need to map these communities back to the batch that contains the contract. For a community $U_j$, we need the batch $b_j$ as the set of unique contracts containing any functions and variables in $U_j$:
\begin{equation}
    b_j = \{ \text{Contract}(f) \mid f \in U_j \}
\end{equation}
This graph-theoretic approach groups functions that interact densely, regardless of file location, often identifying distinct business domains like Tokenomics or Governance. 
The resulting batches $b_j$ represent functional clusters rather than arbitrary file directories, ensuring that the LLM analyzes logically cohesive units.

\noindent
\textbf{Step 3: Graph-Oriented Importance Analysis.}
Within each batch $b_j$, treating all contracts as equally important would dilute the auditor's focus. 
To allocate limited reasoning capacity effectively, we must quantitatively distinguish between core business logic and peripheral components. Therefore, our goal is to identify ``Anchor Contracts'', which are the structural backbones of the batch, by analyzing the obtained subgraph $\mathcal{G}_{b_j}$.

For a contract $c$, its importance score is calculated from the importance of the functions it contains:
\begin{equation}
\label{eq: imp_jc}
\mathrm{Score}_{b_j}(c) = \frac{1}{|c|}\sum_{f \in c} \left( \alpha \cdot \text{Betw}(f) + \beta \cdot \text{PR}(f) \right)
\end{equation}
, where $\text{Betw}(f)$ and $\text{PR}(f)$ denote the \textit{betweenness centrality}~\cite{newman2005measure} and the \textit{page rank}~\cite{pageRankalgorithm} of the given function, respectively, reflecting their importance within $\mathcal{G}_{b_j}$.
Within each batch, contracts are sorted according to their scores, which affects the attention allocation of LLMs~\cite{llmAttention}.

\noindent
\textbf{Step 4: LLM-Driven Refinement.}
While graph clustering captures structural coupling, it may miss semantic business relationships that do not reflect in control or data flow dependencies.
For example, two contracts each call the same library, which forms a batch with each of the two contracts. However, it would be more reasonable to group all three of them into a single batch.
Therefore, we take advantage of LLMs to refine the obtained batches.

LLMs first focus on the business logic coherence among batches.
For each batch $b$, LLMs assign semantic tags (\eg, ``Swap'') to them.
If two batches (\textit{e.g.,} $b_i$ and $b_j$) are considered by LLMs to have the same or related semantics, then they will be merged, \textit{i.e.,} $b_i \cup b_j$.

Although the previous step merged the relevant batches, we still need to consider the LLMs token limit.
Therefore, for the batch $b$ that exceeds the adopted LLM's token limit, LLMs are asked to remove \textit{the most common, i.e., non-customized, contracts within $b$}. 
For example, an ERC-20 template contract may be considered to be removed first, as its semantics must be possessed by LLMs compared to customized token swapping logic.
This step will continue to loop until the size of $b$ meets the token limit requirement.

\subsection{Model-Agnostic Auditing}
\label{subsec: methodology-diagnosis}
\label{sec:method:m2}
For a {\as} task, we abstract it as a tuple
$$\langle \mathcal{K}, \mathcal{B}, \mathcal{V}_{raw}, \Phi \rangle$$
, where $\mathcal{B}=\{b_1,b_2,…,b_n\}$ represents the set of batches obtained in \S\ref{subsec: M1 methodology-profiler} and $\mathcal{K}$ is the knowledge base obtained from public audit checklists and historical attack reports (detailed in \S\ref{subsubsec: M2 core components}). $\Phi$ processes $\mathcal{B}$ to produce the collective vulnerability hypotheses $\mathcal{V}_{raw}$.
Therefore, the auditing process, \textit{i.e.,} how to obtain $\mathcal{V}_{raw}$, is expressed as:
\begin{equation}
\label{eq: v_raw}
    \mathcal{V}_{raw} = \Phi(\mathcal{B}, \mathcal{K})
\end{equation} 

We design a \textit{\textbf{Plan-Remind-Solve}} agentic workflow, which could be high-level interpreted as:
\begin{equation}
\label{eq: phi B,K}
    \Phi(\mathcal{B}, \mathcal{K}) := \bigcup_{b_i \in \mathcal{B}} f_{solve}(f_{plan}(b_i), f_{remind}(\mathcal{K}))
\end{equation}
where $f_{plan}$ identifies potentially vulnerable functions and contracts from the specific batch $b_i$, and $f_{solve}$ conducts the auditing with the help of LLM with the extra knowledge extracted and retrieved by $f_{remind}(\mathcal{K})$.
The complete orchestration of the auditing pipeline is detailed in \S\ref{sec:m2:pipeline}.

\subsubsection{Core Components}
\label{subsubsec: M2 core components}
As stated above, the auditing task relies on three vital objects, \textit{i.e.,} $\mathcal{K}$, $\mathcal{B}$, and $\mathcal{V}_{raw}$. 
As $\mathcal{B}$ has been detailed in \S\ref{subsec: M1 methodology-profiler}, we explain the other two here.

\noindent
\textbf{Knowledge Base ($\mathcal{K}$).} 
We implement $\mathcal{K}$ as a hierarchical file system containing manually curated data from public audit checklists (\eg, Solodit~\cite{auditchecklist}) and historical attack reports posted by prestigious and well-known Web3 security companies/organizations, \eg, DeFiHackLabs~\cite{defihacklabsIncidentExplore} and BlockSec~\cite{blocksecIncidents}. 
Data is classified into distinct vulnerability categories (\eg, reentrancy and donation attack). Each category is stored as a file comprising three segments: \textit{i.e.,} \textit{vulnerability pattern}, \textit{concrete exploit instance}, and \textit{reasoning trace} that capture how expert auditors identify the existence of this vulnerability.
We underline that $\mathcal{K}$ is scalable by updating it to include emerging attack techniques against new vulnerabilities, which ensures the system remains effective against novel threats.

\noindent
\textbf{Vulnerability Hypotheses ($\mathcal{V}_{raw}$).}
$\mathcal{V}_{raw}$ represents the set of identified vulnerability hypotheses prior to verification.
Formally, each hypothesis $v \in \mathcal{V}_{raw}$ is defined as a tuple $\langle e, c, p \rangle$. Here, $e$ denotes the specific \textit{entry point} (function and line number) within a batch; $c$ represents the \textit{constraint set} (preconditions) required for exploitation; and $p$ captures the \textit{reasoning path} derived by matching code features against \textit{reasoning traces} in $\mathcal{K}$.
This structured representation transforms vague suspicions into precise and falsifiable claims, ensuring they are well-defined for the subsequent verification phase.

\newcommand{\RightEq}[1]{\hfill \makebox[0pt][r]{#1}}

\begin{algorithm}[t]
\caption{Model-Agnostic Auditing Pipeline ($\Phi$)}
\label{alg:auditing_pipeline}
\begin{algorithmic}[1]
\REQUIRE Batch $b_i$; Knowledge base $\mathcal{K}$
\ENSURE Raw vulnerability hypotheses $\mathcal{V}_{raw}$

    \STATE \textcolor{blue}{// Step 1: Auditing Task Planning}
    \FORALL{$c_j\ \text{in}\ b_i$}
        \FORALL{$f_{ij}\ \text{in}\ c_j$}
            \IF{$\operatorname{LLMIsVul}(f_{ij}, c_j)$}
                \STATE $\mathcal{F}_{ij} \leftarrow \operatorname{ExtractCallerCallee}(f_{ij}, \mathcal{G})$
                \STATE $\mathcal{K}_{ij} \leftarrow \operatorname{LLMRelate}(\mathcal{F}_{ij}, \mathcal{K})$
                \STATE $\mathcal{Q}.\operatorname{append}((\mathcal{F}_{ij}, \mathcal{K}_{ij}, c_j))$
            \ENDIF
        \ENDFOR
    \ENDFOR
    \STATE $\mathcal{Q}.\operatorname{sort}(\operatorname{key}=(Score_{b_i}(q.c), \operatorname{LLMBySeverity}(q.\mathcal{F})))$
    
    \FORALL{$q\ \text{in}\ \mathcal{Q}$} 
        \STATE \textcolor{blue}{// Step 2: Remind \& Knowledge Augment}
        \STATE $K_{context} \leftarrow q.\mathcal{K}$
        \IF{$\text{MCPIsDerivative}(q.c)$}
            \STATE $K_{context}.\operatorname{update}(\operatorname{MCPSearch}(q.c))$
        \ENDIF
    
        \STATE \textcolor{blue}{// Step 3: Solve}
        \STATE $\mathcal{V}_{base} \leftarrow \operatorname{LLMAudit}(b_i, q.\mathcal{F}, K_{context})$
    
        \STATE \textcolor{blue}{// Step 3-1: Adversarial State Assumption Analysis}
        \STATE $\mathcal{V}_{adv} \leftarrow \emptyset$
        \FORALL{$p\ \text{in}\ \mathcal{P}_{adv}$}
            \STATE $\mathcal{V}_{adv}.\operatorname{update}(\operatorname{LLMAudit}(b_i, q.\mathcal{F}, K_{context} + p))$
        \ENDFOR
    
        \STATE \textcolor{blue}{// Step 3-2: Targeted Symbolic Constraint Solving}
        \STATE $\mathcal{F}_{sen\_ari} \leftarrow \operatorname{LLMSensitiveArith}(q.\mathcal{F}), \mathcal{V}_{math}\leftarrow\emptyset$
        \FORALL{$f\ \text{in}\ \mathcal{F}_{sen\_ari}$}
            \STATE $K_{ari} \leftarrow \operatorname{Z3Solve}(\operatorname{LLMTranspile}(f)) + \pi_{real}$
            \STATE $\mathcal{V}_{math}.\operatorname{update}(\operatorname{LLMAudit}(b_i, \operatorname{ExtractCallerCallee}(f, \mathcal{G}), K_{ari}))$
        \ENDFOR
    \ENDFOR

    \STATE \textcolor{blue}{// Step 3-3: Global Synthesis via Iterative Combination}
    \STATE $\mathcal{V}_{raw} \leftarrow \mathcal{V}_{base} \cup \mathcal{V}_{adv} \cup \mathcal{V}_{math}$
    \REPEAT
        \STATE $\text{merged} \leftarrow \text{FALSE}$
        \FORALL{$\{v_i, v_j\}\ \text{in}\ \mathcal{V}_{raw}$}
            \IF{$\operatorname{LLMisLink}(v_i, v_j)$}
                \STATE $\mathcal{V}_{raw} \leftarrow (\mathcal{V}_{raw} \setminus \{v_i, v_j\}) \cup \{ \operatorname{Chain}(v_i, v_j) \}$
                \STATE $\text{merged} \leftarrow \text{TRUE}$
                \STATE \textbf{break} 
            \ENDIF
        \ENDFOR
    \UNTIL{not \text{merged}}

\STATE \textbf{return} $\mathcal{V}_{raw}$
\end{algorithmic}
\end{algorithm}

\subsubsection{Auditing Pipeline}
\label{sec:m2:pipeline}
As formally defined, the auditing pipeline executes the \textit{Plan-Remind-Solve} workflow iteratively for each code batch $b_i \in \mathcal{B}$. We structurally decompose this process into three distinct steps, as detailed in Algorithm~\ref{alg:auditing_pipeline}.

\noindent
\textbf{Step 1: Plan (L1--L8\footnote{The step corresponds to Line 1 to Line 8 in Algorithm~\ref{alg:auditing_pipeline}. We adopt such notations in the following.})}
The process initiates with the \textit{Plan} phase ($f_{plan}$). This step receives a code batch $b_i$ and its corresponding function dependency graph $\mathcal{G}$, and tries to formulate an auditing task queue $\mathcal{Q}$:
\begin{equation}
\label{eq: Q:=FK}
\small
    \mathcal{Q} := [(\mathcal{F}_1, \mathcal{K}_1, c_1), (\mathcal{F}_2, \mathcal{K}_2, c_2), \dots, (\mathcal{F}_m, \mathcal{K}_m, c_m)]
\end{equation}
, where for each task $(\mathcal{F}_i, \mathcal{K}_i, c_i)$, $\mathcal{F}_i$ is a set of functions extracted from $\mathcal{G}$, $c_i$ refers to the corresponding contract, and $\mathcal{K}_i$ represents the knowledge relevant to these functions.

Specifically, within each $b_i$, by iterating all contracts, LLMs are asked to identify \textit{are there functions vulnerable to possible attacks.}\footnote{The prompt word fragment is shown in Appendix~\ref{appendix:prompt1}}
Based on $\mathcal{G}$, callers and callees of these identified functions could be extracted to form $\mathcal{F}_i$.
Then, for each $\mathcal{F}_i$, LLMs are asked to extract related files from $\mathcal{K}$ as the customized knowledge base for the following auditing phase.\footnote{The prompt word fragment is shown in Appendix~\ref{appendix:prompt2}}
We emphasize that $\mathcal{Q}$ is sorted first based on the contract's score (see Eq.~\ref{eq: imp_jc}), and secondly based on the potential serious consequences of the identified function.\footnote{The prompt word fragment is shown in Appendix~\ref{appendix:prompt3}}
This sorting maximizes the ability of subsequent auditing to focus on the more important auditing tasks at the beginning.

\noindent
\textbf{Step 2: Remind (L10 -- L13).}
For each auditing task, the \textit{Remind} process retrieves relevant and concrete knowledge for the given contract $c$ from two sources, \textit{i.e.,} constructed knowledge base $\mathcal{K}$ and via MCP servers.

Specifically, it first constructs the knowledge context $K_{context}$ by retrieving a specific exploit instance and reasoning trace for the given contract. $K_{context}$ serves as a cognitive trigger, guiding the agent to reflect on edge cases relevant to the task.
However, the static nature of $\mathcal{K}$ may miss emerging vulnerabilities. 
To mitigate this, we integrate a custom Model Context Protocol (MCP) server for real-time augmentation. We utilize an MCP tool \texttt{check\_protocol\_lineage} to analyze if the contract is a derivative of known DeFi primitives (\eg, a fork of Compound). If it is, it triggers another MCP tool to execute live search queries about the newly happened incidents with that DeFi protocol. 
The search results are parsed into a summarized markdown format and merged into $K_{context}$.

\noindent
\textbf{Step 3: Solve (L14 -- L33).}
Finally, the \textit{Solve} phase is going to identify business logic vulnerabilities for each task, \textit{i.e.,} generate vulnerability hypotheses.
{\tool} first conducts a basic audit with the help of LLMs. Specifically, given a batch $b_i$, taking the functions in $q.\mathcal{F}$ as potential entries, LLMs are asked to perform auditing based on knowledge $K_{context}$, formally represented as:
\begin{equation}
\label{eq: v_base}
    \mathcal{V}_{base} = \operatorname{LLMAudit}(b_i, q.\mathcal{F}, K_{context})
\end{equation}
However, this process is similar to existing LLM-based work~\cite{sun2024gptscan}, which has been shown to suffer from severe false positive and negative issues.
Therefore, taking a step forward, we enhance the capabilities of this phase by integrating the following three additional auditing tasks: \textit{adversarial state assumption analysis}, \textit{targeted symbolic constraint solving}, and \textit{global synthesis via iterative combination}.

\textit{(1) Adversarial State Assumption Analysis (L16 -- L19).}
Beyond the basic auditing, to capture vulnerabilities that only manifest under extreme hostile conditions, we propose an \textit{adversarial state injection} mechanism. 
Specifically, we iterate the solver over a set of distinct adversarial profiles $\mathcal{P}_{adv}$, where each profile represents a specific class of state tampering, formally represented as:
\begin{equation}
\label{eq: v_role}
    \mathcal{V}_{adv} = \bigcup_{p \in \mathcal{P}_{adv}} \operatorname{LLMAudit}(b_i, q.\mathcal{F}, K_{context}+p)
\end{equation}
The $p$ refers to the reasoning context to enforce aggressive, malicious assumptions. We categorize these profiles $\mathcal{P}_{adv}$ into three dimensions:
\begin{itemize}[leftmargin=*]
    \item \textbf{Environment Tampering:} Assuming the full control over blockchain variables (which is reasonable for miners), like manipulating \texttt{block.timestamp} to trigger time-lock bypasses or manipulating \texttt{block.number} to exploit pseudo-randomness.
    \item \textbf{Interaction Hijacking:} Assuming the full control over external call entities, including injecting malicious \texttt{calldata} to trigger edge cases and forcing external calls to return manipulated values (\eg, returning \texttt{false} on success).
    \item \textbf{Resource Infinity:} Assuming the ability to manipulate a significant amount of tokens, like utilizing Flash Loans to acquire almost infinite capital for exploiting price manipulation that is invisible under normal balance assumptions.
\end{itemize}

\textit{(2) Targeted Symbolic Constraint Solving (L20 -- L24).}
While LLMs excel at semantic logic, detecting arithmetic precision loss is notoriously difficult for neural models. To balance the efficiency and rigor, {\tool} takes the formal constraint solving results as part of the knowledge:
\begin{equation}
\label{eq: v_math}
\small
\begin{split}
    \mathcal{V}_{math} =\!\!\!\!\!\! \bigcup_{f \in \mathcal{F}_{sen\_ari}}\!\!\!\!\!\! &\operatorname{LLMAudit}(b_i, \mathcal{F}, \operatorname{Z3Solve}(\operatorname{LLMTranspile}(f))\! +\! \pi_{real})\\
    &\text{, where}\ \mathcal{F} = \operatorname{ExtractCallerCallee}(f, \mathcal{G})
\end{split}
\end{equation}
, where $\mathcal{F}_{sen\_ari} \subset q.\mathcal{F}$, denoting functions with sensitive numerical operations (\eg, division before multiplication, mixed-unit conversion).
The function $\text{LLMTranspile}(\cdot)$ leverages LLMs to convert the Solidity statements in $\mathcal{F}_{sen\_ari}$ into Z3 statements, mapping EVM data types (like \texttt{uint256}) to SMT bit-vectors.
Beyond constraints derived from functions, like \texttt{require} statements, we explicitly add a set of constraints $\pi_{real}$ to avoid obtaining unrealistic values, such as negative balance.
The function $\text{Z3Solve}$ then attempts to find a set of concrete solutions for the above constraint set.

\textit{(3) Global Synthesis via Iterative Combination (L25 -- L34).}
After processing all tasks in $\mathcal{Q}$ and executing symbolic checks, we then try to address the \textit{cognitive tunneling} limitation~\cite{wei2024llmaudit}, \textit{i.e.,} the model's tendency to focus on individual, isolated vulnerabilities while failing to recognize potential multi-step exploit chains that require connecting distinct logic flows, by an \textit{iterative combination analysis} to derive $\mathcal{V}_{syn}$, which is formally expressed as:
\begin{equation}
\label{eq: v_compo}
\begin{split} 
    \mathcal{V}_{syn}^{(t+1)} &= \mathcal{V}_{syn}^{(t)} \cup \{ \text{Chain}(v_i, v_j)\}\\
    &\text{, where } v_i, v_j \in \mathcal{V}_{raw} \cup \mathcal{V}_{syn}^{(t)}\ \text{and}\ \operatorname{LLMisLink}(v_i, v_j)
\end{split}
\end{equation}
, where $\operatorname{LLMisLink}(v_i, v_j)$ determines if the post-condition of $v_i$ satisfies the pre-condition of another one $v_j$. 
If a valid path exists, $\text{Chain}$ constructs a composite exploit scenario, \textit{i.e.,} taking these two as a single vulnerability hypothesis. 
This process iterates until no new chains between vulnerabilities are found. Consequently, it guides the model to conceptualize complex financial attack scenarios formed by chaining isolated bugs across the entire batch

\subsection{False Positive Filtration}
\label{subsec: methodology-verify}
\label{sec:method:m3}
To minimize false positives and pursue high-precision reporting, we introduce a dedicated verification phase, denoted as $\Psi$. This pipeline addresses the primary limitations of the auditing phase (specifically context gaps, redundancy, and unrealistic assumptions) by processing the raw hypothesis set $\mathcal{V}_{raw}$ through three sequential logic filters.
We formally define the verification process as a composite function:
\begin{equation}
\small
    \mathcal{V}_{final} = \Psi(\mathcal{V}_{raw}) := \text{map}(f_{feas}\cdot f_{dedup}\cdot f_{agg}, \mathcal{V}_{raw})
\end{equation}
The complete orchestration of the proposed verification pipeline is detailed in the following.

\noindent
\textbf{Step 1: Contextual Aggregation ($f_{agg}$).}
The previous auditing phase may suffer from false positives due to incomplete context. 
For example, \texttt{bar} is vulnerable to the reentrancy vulnerability, which is invoked by \texttt{foo}. However, \texttt{foo} is sanitized by the \texttt{nonReentrant} modifier. 
To resolve this, given a $v \in \mathcal{V}_{raw}$, {\tool} is asked to reevaluate the feasibility of $v$ with the help of LLMs.
Recall $v$ is composed of entry function $e$, constraint set $c$, and reasoning path $p$ (see \S\ref{subsubsec: M2 core components}). Therefore, {\tool} is asked to reevaluate if, following $p$ starting from $e$, the jump conditions among steps in $p$ meet $c$.
By evaluating $p$ under a global context, this step effectively filters out local risks that are neutralized by global safeguards.

\noindent
\textbf{Step 2: Semantic Deduplication ($f_{dedup}$).}
Raw auditing frequently generates redundant reports where distinct entry points trigger the same underlying logic error. To address this, we employ a semantic clustering approach to map hypotheses to their root causes and merge duplicates.

Specifically, for each $v$, we obtain its embedding: $\mathbf{v}_i = f_{emb}(v_i)$. Then, we adopt the cluster method on embeddings to obtain a set of clusters, \textit{i.e.,} ${C_1, C_2, \dots, C_k}$.
For each cluster, we intend to retain the one with the highest confidence score evaluated by LLMs.
Formally, 
\begin{equation}
\label{eq: v_uniq}
\small
\begin{split}
    \mathcal{V}_{uniq} &:= f_{dedup}(\mathcal{V}_{raw}) = \{v^*\}, \text{where}\ v_i \in \mathcal{V}_{raw} \ \text{and}\ \\
    v^* &= \operatorname{selectBest}(C_j, \lambda_{\text{LLM}}) = \underset{v \in \{v_i \mid \mathbf{v}_i \in C_j\}}{\operatorname{arg\,max}} \, \lambda_{\text{LLM}}(v)
\end{split}
\end{equation}
Crucially, we attach the reasoning path $p$ of other members in the same cluster to this canonical entry as ``alternative trigger paths'', ensuring that while report volume is reduced, no diagnostic information is lost.

\noindent
\textbf{Step 3: Threat Model Assessment ($f_{feas}$).}
We underline that the previous auditing phase may produce hypotheses based on external threats. For example, LLMs may assume dishonest developers (\textit{e.g.,} obtaining admin privileges with backdoors) and unreliable infrastructures (\textit{e.g.,} misbehaving oracles).
To this end, $f_{feas}$ evaluates if the assumption of exploiting $v$ requires exploiting external entities, which is deemed technically infeasible and discarded.

\begin{figure*}[t]
    \centering
    \includegraphics[width=\linewidth]{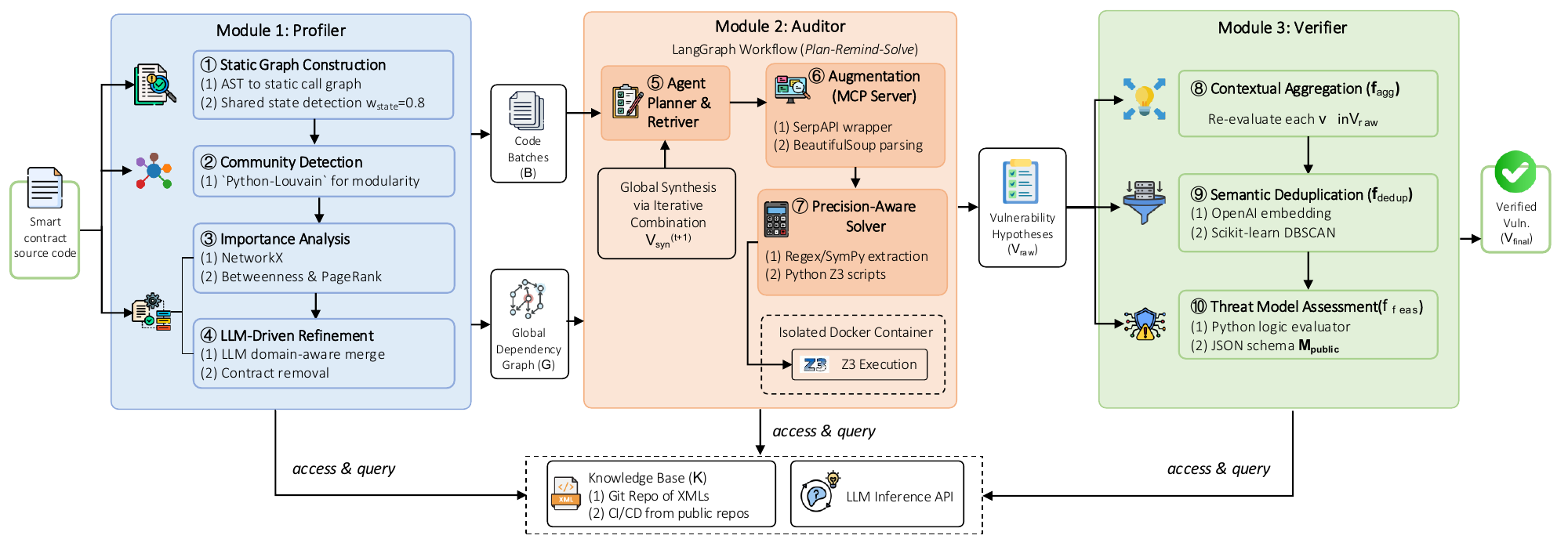}
    \caption{Implementation of {\tool}}
    \label{fig:implementation}
\end{figure*}

\section{Implementation}
\label{sec:implementation}
We implement {\tool} as a modular framework consisting of over 7K lines of Python3 code. {\tool} follows a modular design principle, corresponding to our methodology introduced from \S\ref{sec:method:m1} to \S\ref{sec:method:m3}, \textit{i.e.,} \textit{Profiler}, \textit{Auditor}, and \textit{Verifier}, respectively. As shown in Fig.~\ref{fig:implementation}, they operate sequentially: \textit{Profiler} provides the structural foundation ($\mathcal{G}$ and $\mathcal{B}$); \textit{Auditor} generates the candidate vulnerability hypotheses ($\mathcal{V}_{raw}$); and \textit{Verifier} executes the filtration process to obtain $\mathcal{V}_{final}$.
All modules share access to the underlying LLM inference API and the local Knowledge Base $\mathcal{K}$.

\subsection{Profiler}
\label{subsec: implementation:profiler}
Profiler takes raw smart contract source code files as input and outputs context-preserving code batches $\mathcal{B}$, along with a global function dependency graph $\mathcal{G}$.
To construct $\mathcal{G}$ (Step 1 in \S\ref{sec:method:m1}), we utilize Slither~\cite{slither} to parse the Abstract Syntax Tree (AST) and generate the initial static call graph. To capture data flow dependence relationships among functions, we extend Slither's API to detect shared state variable access.
To avoid rebuilding the wheel, both the community detection (Step 2 in \S\ref{sec:method:m1}) and the importance analysis (Step 3 in \S\ref{sec:method:m1}) are built on existing libraries, \textit{i.e.,} python-louvain~\cite{louvain} and NetworkX~\cite{networkX}, respectively. 
Regarding the refinement process (Step 4 in \S\ref{sec:method:m1}), LLMs are asked to output a list of predefined semantic tags (\eg, ``Governance'', ``Liquidity'') for each batch. A Python script then parses these tags and automatically merges batches sharing identical high-level semantic tags.
To enforce the token limit ($L_{max}=32k$), we implement an iterative pruning loop. Instead of arbitrary truncation, the system queries the LLM to identify and remove standardized library contracts (\eg, OpenZeppelin's libraries) from the overflowed batch. This prioritization logic assumes that template code is already internalized by the model's pre-training, allowing developers to preserve custom business logic while satisfying context constraints.

\subsection{Auditor}
\label{subsec: implementation:auditor}
Auditor takes the batch set $\mathcal{B}$, the dependency graph $\mathcal{G}$, and the Knowledge Base $\mathcal{K}$ as input to produce vulnerability hypotheses $\mathcal{V}_{raw}$.
Following the ``Plan-Remind-Solve'' workflow described in \S\ref{sec:method:m2}, Auditor is implemented with the help of LangGraph~\cite{langgraph}. 

Specifically, $\mathcal{K}$ is maintained as a Git repository of XML-formatted vulnerability patterns. A CI/CD pipeline automatically parses new Markdown reports from public repositories into this structured format.
To achieve real-time online knowledge augmentation, we deployed a custom MCP server wrapping the SerpAPI~\cite{serpapi}, which uses BeautifulSoup to parse search results and injects derivative protocol vulnerabilities directly into the agent's context window.
Regarding the solve phase in Auditor, it implements the three-stage auditing strategies introduced in \S\ref{sec:method:m2}. 
For the targeted symbolic constraint solving, we implemented a transpiler using Regex and SymPy to extract arithmetic operations and convert them into Python Z3 scripts with the help of LLMs. These scripts are executed in an isolated Docker container to verify satisfiability without risking host system stability.

\subsection{Verifier}
\label{subsec: implementation:verifier}
Verifier takes raw hypotheses $\mathcal{V}_{raw}$ as input and outputs the final verified set $\mathcal{V}_{final}$. It implements the three-stage filtering logic introduced in \S\ref{sec:method:m3}.
To conduct semantic deduplication ($f_{dedup}$, Step 2 in \S\ref{sec:method:m3}), we use OpenAI's text-embedding-3-small to generate 1536-dimensional embeddings. Clustering is performed via Scikit-learn's DBSCAN implementation ($\epsilon=0.15, \text{min\_samples}=1$).

\section{Evaluation}
\label{sec:evaluation}
In this section, we present the experimental design and all evaluation results.

\subsection{Experimental Design}

\subsubsection{Adopted LLMs}
\label{subsec: underlying models}
As stated in \S\ref{sec:method}, {\tool} relies on LLMs to achieve the auditing task.
This part introduces the selection and configuration criteria of adopted LLMs in the evaluation.

\noindent
\textbf{LLM Selection.}
To illustrate the model agnosticism of {\tool}, for a fair comparison, we select a diverse array of LLMs, including \textit{high-capacity proprietary LLMs} (\eg, Gemini-2.5-Pro (128B) and Claude-Sonnet-4.5), \textit{cost-effective proprietary LLMs} (\eg, Gemini-2.5-Flash (5B) and Claude-Haiku-4.5 (20B)), and \textit{lightweight open-source LLMs} (\eg, GPT-oss-20B (20B) and GPT-oss-120B (120B)). 

\noindent \textbf{LLM Configuration.}
We configure the \textit{temperature} at 0.7 and \textit{top-p} at 0.9. While this setting promotes exploratory reasoning to uncover obscure bugs, it introduces potential nondeterminism. 
To validate the reliability of using single-execution results, we conducted a stability pilot study involving three independent audit iterations per model on a random project.
Specifically, we measured the consistency using Sentence-BERT~\cite{reimers2019sentence} for semantic similarity and the Friedman test~\cite{F_T} for variance analysis.
The results demonstrate high stability, particularly for severe vulnerabilities: while minor findings showed slight variance, the semantic consistency for Critical and High-severity vulnerabilities exceeded \textbf{0.95} across iterations. The Friedman test yielded a p-value of 0.56, indicating no statistically significant divergence in the overall audit quality.
This confirms that {\tool} reliably reproduces critical security insights even under exploratory parameters, justifying our adoption of single-execution evaluation.

\subsubsection{Datasets \& Baseline}
We introduce how we construct datasets and select baselines for a fair and comprehensive evaluation for {\tool}.

\noindent \textbf{Datasets.}
We have constructed three distinct datasets.

\begin{itemize}[leftmargin=*]
    \item \textbf{Dataset 1: Real-world High-value Exploits ($\mathcal{D}_1$).} 
    This dataset comprises the top 20 attacks that have caused the largest financial losses since June 2025. The ground truth consists of public analysis reports and Proof of Concepts (PoCs) for these exploits. Since the latest LLM we adopted is Gemini-2.5-Pro~\cite{gemini2.5pro}, which was released by the vendor after June 2025, this temporal separation guarantees that \ul{none of the adopted LLMs included these specific incidents during their training phases}. Consequently, this evaluation tests {\tool}'s actual reasoning capabilities rather than its ability to recall training data.

    \item \textbf{Dataset 2: Proprietary Benchmark ($\mathcal{D}_2$).} 
    This dataset comprises the top 30 exploit events from the DeFi exploit benchmark published by Anthropic~\cite{benchmarkAnthropic}. Similar to $\mathcal{D}_1$, the ground truth consists of public analysis reports and PoCs. We utilize this dataset to facilitate fair comparisons, allowing us to benchmark {\tool} directly against existing state-of-the-art tools, \textit{i.e.,} the Claude auditing agent~\cite{benchmarkAnthropic}.

    \item \textbf{Dataset 3: Crowdsourced Auditing Projects ($\mathcal{D}_3$).} 
    This dataset includes the top 30 highest-rewarding contest projects from Sherlock~\cite{sherlockContest} since the beginning of 2025. The ground truth consists of the final auditing reports from these contests. As Sherlock contests allow participation from all auditors, this crowdsourced approach results in highly comprehensive vulnerability reports. Any finding by {\tool} that is not present in the contest report will be initially classified as a potential false positive. 
\end{itemize}

\noindent \textbf{Baselines.}
We selected four representative tools from both industry and academia.
Specifically, Claude-family LLMs represent the leading industry standard for LLM-based auditing. We directly cite the performance metrics reported in Anthropic's official technical paper~\cite{benchmarkAnthropic}.
GPTScan~\cite{xia2024auditgpt} and LLMSmartAudit~\cite{wei2024llmaudit} are academic tools. Among the academic prototypes proposed recently, these are the only two open-source tools that we successfully deployed and executed in our local environment, ensuring reproducible baselines.
Hound~\cite{hound} is a cutting-edge open-source auditing framework widely used in the industry, representing the best available non-academic static analysis tool.

\subsubsection{Research Questions}
To comprehensively evaluate {\tool}, we plan to answer the following research questions (RQs):

\begin{itemize}[leftmargin=*]
    \item \textbf{RQ1 (Effectiveness): How effective is {\tool} when auditing smart contracts?}
    \begin{itemize}[leftmargin=*]
        \item (RQ1.1) \textit{Can {\tool} identify intricate complex business-logic vulnerabilities?}
        \item (RQ1.2) \textit{Can {\tool} effectively reduce the false positive rate?}
    \end{itemize}

    \item \textbf{RQ2 (Cost-Efficiency): What is the cost-efficiency of {\tool} in terms of time and monetary resources?}

    \item \textbf{RQ3 (Ablation Study): What is the indispensability of the Verifier in {\tool}?} 

    \item \textbf{RQ4 (Case Study): Can {\tool} identify previously unknown vulnerabilities in the wild?}
\end{itemize}

To systematically address these proposed research questions, we designed specific evaluation protocols tailored to different datasets and baselines.

For RQ1, we evaluate {\tool}'s effectiveness from two dimensions: \textit{exploit detection capability} (RQ1.1) and \textit{false positive reduction} (RQ1.2). 
To answer RQ1.1, we execute {\tool} and all baselines on $\mathcal{D}_1$ and $\mathcal{D}_2$. Since a binary ``found/not found'' metric lacks nuance, we implement an LLM-based judge based on Claude-Sonnet-4.5 to semantically compare the generated audit reports against the ground truth, \ie, released post-mortem report. The judge evaluates consistency across three aspects: \textit{root cause}, \textit{entry point}, and \textit{attack path}. To be classified as a successful detection, a finding must, at a minimum, identify the correct vulnerability type or location. Reports providing a more comprehensive analysis, including the root cause and attack path, are naturally included in this count.
To answer RQ1.2, we benchmark {\tool} on $\mathcal{D}_3$ using a customized version of ScaBench~\cite{scabench}, which automates the alignment between {\tool}'s findings and human audit reports based on code location and vulnerability category. We measure \textit{detection rate}, \textit{precision}, and \textit{f1-score}. Crucially, for any false positives reported by {\tool} but absent in the contest report, we perform a manual verification. If confirmed as a valid security issue missed by the contest participants, it is reclassified as a true positive. 

To answer RQ2, during the evaluation of RQ1, we also record the number of tokens and the time spent auditing each project and normalize these into monetary costs based on the API pricing at the time of writing. By correlating \textit{detection rate}, \textit{precision}, and \textit{f1-score} with the cost metrics, we analyze the trade-off between performance and affordability.

For RQ3, we create a variant of {\tool}: \textit{without Verifier}, denoted as ${\tool}_{w/o-V}$ (skipping the verification). We compare the \textit{detection rate}, \textit{precision}, and \textit{f1-score} of ${\tool}_{w/o-V}$ against {\tool} on $\mathcal{D}_3$.

For RQ4, we evaluate {\tool}'s practical utility through real-world deployments, including auditing a \$400M protocol, collaborating with firms, and participating in public contests. We analyze confirmed zero-day vulnerabilities found in the wild and retrospectively assess {\tool}'s ability to reproduce recent high-value exploits

\subsubsection{Environmental Setup.}
All experiments were conducted on a Mac Studio workstation equipped with an Apple M3 Ultra chip (32-core CPU, 32-core GPU) and 512GB of Unified Memory.
For proprietary LLMs, we accessed the inference services via the OpenRouter API~\cite{openrouterModel} to ensure standardized latency and rate limits. Also, the monetary cost is calculated based on OpenRouter's API price.
For open-source models, specifically GPT-oss-20B and GPT-oss-120B, we deployed them locally on the workstation using the Ollama runtime to minimize network overhead.
{\tool} leverages langgraph 0.6.7 for building the agentic workflow.

\subsection{RQ1: Effectiveness}

\begin{table}[t]
  \centering
  \caption{Overall performance of {\tool} and baselines on $\mathcal{D}_1$ and $\mathcal{D}_2$, where C, G, L, H refer to Claude(s), GPTScan, LLMSmartAudit, and Hound, respectively.}
    \resizebox{\linewidth}{!}{
    \begin{tabular}{lccccc}
    \toprule
          & \textbf{C}$^1$ & \textbf{G} & \textbf{L} & \textbf{H} & \textbf{{\tool}}$^2$ \\
    \midrule
    \textbf{Detection Rate} & {51.1\%} & {5.5\%} & {9.7\%} & {20.0\%} & \textbf{86.7\%}$^{***}$ \\ \midrule
    \textbf{avg.$^3$ Monetary Cost} & \$1.22 & \$0.69 & \$0.85 & \$17.80 & \textbf{\$0.22} \\
    \textbf{avg.$^3$ Time Cost} & -     & 31s & 795s & 7,488s & \textbf{190s} \\
    \bottomrule
    \multicolumn{6}{l}{\begin{tabular}[c]{@{}l@{}}$^1$ Data of Claude(s) only represents its reported results on $\mathcal{D}_2$.\\ $^2$ Data of {\tool} is with Claude-Sonnet-4.5 as the LLM. \\ $^3$ The avg. is the average cost per project audit, aligned with\\the reporting metric in the Claude benchmark. \end{tabular}}\\
    \end{tabular}
}
\label{tab:performance on D1, D2}
\end{table}%

\begin{table}[t]
  \centering
  \caption{The performance of baselines on $\mathcal{D}_3$, where G, L, H refer to GPTScan, LLMSmartAudit, and Hound, respectively.}
    \begin{tabular}{lccc}
    \toprule
     & \textbf{G} & \textbf{L} & \textbf{H} \\
    \midrule
    \textbf{Detection Rate}  & {1.37\%} & {31.36\%} & {0.64\%} \\
    \textbf{False Negative Rate} & {98.63\%} & {68.64\%} & {99.36\%} \\
    \textbf{False Positive Rate}   & {97.50\%} & {99.01\%} & {80.20\%} \\
    \textbf{Precision}  & {2.50\%} & {0.99\%} & {19.80\%} \\
    \textbf{F1 Score} & {\cellcolor[rgb]{ .851,  .851,  .851}0.0177} & {\cellcolor[rgb]{ .851,  .851,  .851}0.0192} & {\cellcolor[rgb]{ .851,  .851,  .851}0.0124} \\
    \midrule
  \textbf{avg.$^1$ Monetary Cost} & \$8.25 & \$1.26 & \$9.60 \\
    \textbf{avg.$^1$ Time Cost}  & 293s & 7,224s & 27,325s \\
    \bottomrule
    \multicolumn{4}{l}{\begin{tabular}[c]{@{}l@{}}$^1$ The average cost per 10K LOC in smart contracts. \end{tabular}}\\
    \end{tabular}%
  \label{tab:baseline-perf}%
\end{table}%

\begin{table*}[t]
  \centering
  \caption{The performance of {\tool} and ${\tool}_{w/o-V}$ on $\mathcal{D}_3$ with different underlying LLMs.}
  \resizebox{\linewidth}{!}{
    \begin{tabular}{lcccccc}
    \toprule
          & {\textbf{Claude-Haiku-4.5}} & {\textbf{Claude-Sonnet-4.5}} & {\textbf{GPT-oss-20B}} & {\textbf{GPT-oss-120B}} & {\textbf{Gemini-2.5-Flash}} & {\textbf{Gemini-2.5-Pro}} \\
    \midrule
    \textbf{Detection Rate} & 61.69\%$^{***}$ & 78.92\%$^{***}$ & 32.66\%$^{***}$ & 34.78\%$^{***}$ & 57.02\%$^{***}$ & 40.38\%$^{***}$ \\
    \textbf{False Negative Rate$^1$} & 38.31\% & 21.08\% & 67.34\% & 65.22\% & 42.98\% & 59.62\% \\
    \textbf{False Positive Rate$^1$} & 49.12\% & 47.36\% & 31.27\% & 36.60\% & 31.84\% & 36.27\% \\
    \textbf{Precision} & 50.88\% & 52.64\% & 68.73\% & 63.40\% & 68.16\% & 63.73\% \\
    \rowcolor[rgb]{ .851,  .851,  .851} \textbf{F1 Score} & 0.5577$^{*}$ & 0.6316$^{*}$ & 0.4428$^{*}$ & 0.4492$^{*}$ & 0.6209$^{*}$ & 0.4944$^{*}$ \\
    \midrule
    \textbf{avg. Monetary Cost$^2$} & {\$0.76} & {\$2.31} & N/A      & N/A     & {\$0.59} & {\$2.12} \\
    \textbf{avg. Time Cost$^2$} & {657.30s} & {1,429.82s} & {351.45s} & {269.98s} & {984.23s} & {1,033.18s} \\
    \bottomrule
    \toprule
    \textbf{${\tool}_{w/o-V}$} & {\textbf{Claude-Haiku-4.5}} & {\textbf{Claude-Sonnet-4.5}} & {\textbf{GPT-oss-20B}} & {\textbf{GPT-oss-120B}} & {\textbf{Gemini-2.5-Flash}} & {\textbf{Gemini-2.5-Pro}} \\
    \midrule
    \rowcolor[rgb]{ .851,  .851,  .851} \textbf{Detection Rate} & 84.62\% & 92.45\% & 38.46\% & 45.31\% & 69.23\% & 49.97\% \\
    \textbf{False Negative Rate$^1$} & 15.38\% & 7.55\% & 61.54\% & 54.69\% & 30.77\% & 50.03\% \\
    \textbf{False Positive Rate$^1$} & 86.16\% & 76.82\% & 70.60\% & 63.33\% & 68.76\% & 72.27\% \\
    \textbf{Precision} & 13.84\% & 23.18\% & 29.40\% & 36.67\% & 31.24\% & 27.73\% \\
    \textbf{F1 Score} & 0.2379 & 0.3707 & 0.3333 & 0.4053 & 0.4305 & 0.3567 \\
    \bottomrule
    \multicolumn{6}{l}{\begin{tabular}[c]{@{}l@{}} $^1$ False Negative Rate = False Negative / Total Expected; False Positive Rate = False Positive / Total Found.\\ $^2$ The average represents the monetary/time cost of auditing 10K LOC in smart contracts.\\ $^{***}$ indicates statistically extreme significant improvement ($p < 0.001$) compared to baselines.\\ $^{*}$ indicates a significant improvement ($p < 0.05$) compared to ${\tool}_{w/o-V}$. \end{tabular}}\\
    \end{tabular}%
    }
  \label{tab:rq_d3_performance}%
\end{table*}%

This section evaluates the overall effectiveness of {\tool} by examining its ability to detect complex exploits (RQ1.1) and its precision in filtering noise (RQ1.2).

\subsubsection{RQ1.1: Detection of Complex Business Logic Vulnerabilities}
The primary objective of RQ1.1 is to evaluate the ability of {\tool} and all baselines on identifying intricate business-logic vulnerabilities. We aggregated the results of $\mathcal{D}_1$ and $\mathcal{D}_2$ in Table~\ref{tab:performance on D1, D2}.

Powered by Claude-Sonnet-4.5, {\tool} demonstrates exceptional efficacy, achieving a combined detection rate of \textbf{86.7\%}. This performance represents a significant leap over the current state-of-the-art.
Notably, the baseline \textit{Claude(s)}~\cite{benchmarkAnthropic}, despite utilizing the same underlying model family, only achieved 51.1\%. The performance gap serves as the direct evidence that our agentic framework (specifically the \textit{Plan-Remind-Solve} workflow) contributes substantially to detection capability, independent of the model's raw intelligence.
Furthermore, existing tools struggled to capture these complex logic bugs: \textit{Hound} peaked at 20.0\%, while academic prototypes like \textit{GPTScan} (5.5\%) and \textit{LLMSmartAudit} (9.7\%) proved largely ineffective against modern DeFi exploits.

\subsubsection{RQ1.2: Reduction of False Positives}
\begin{figure}
    \begin{minipage}[t]{0.495\linewidth}
        \centering
        \includegraphics[width=\linewidth]{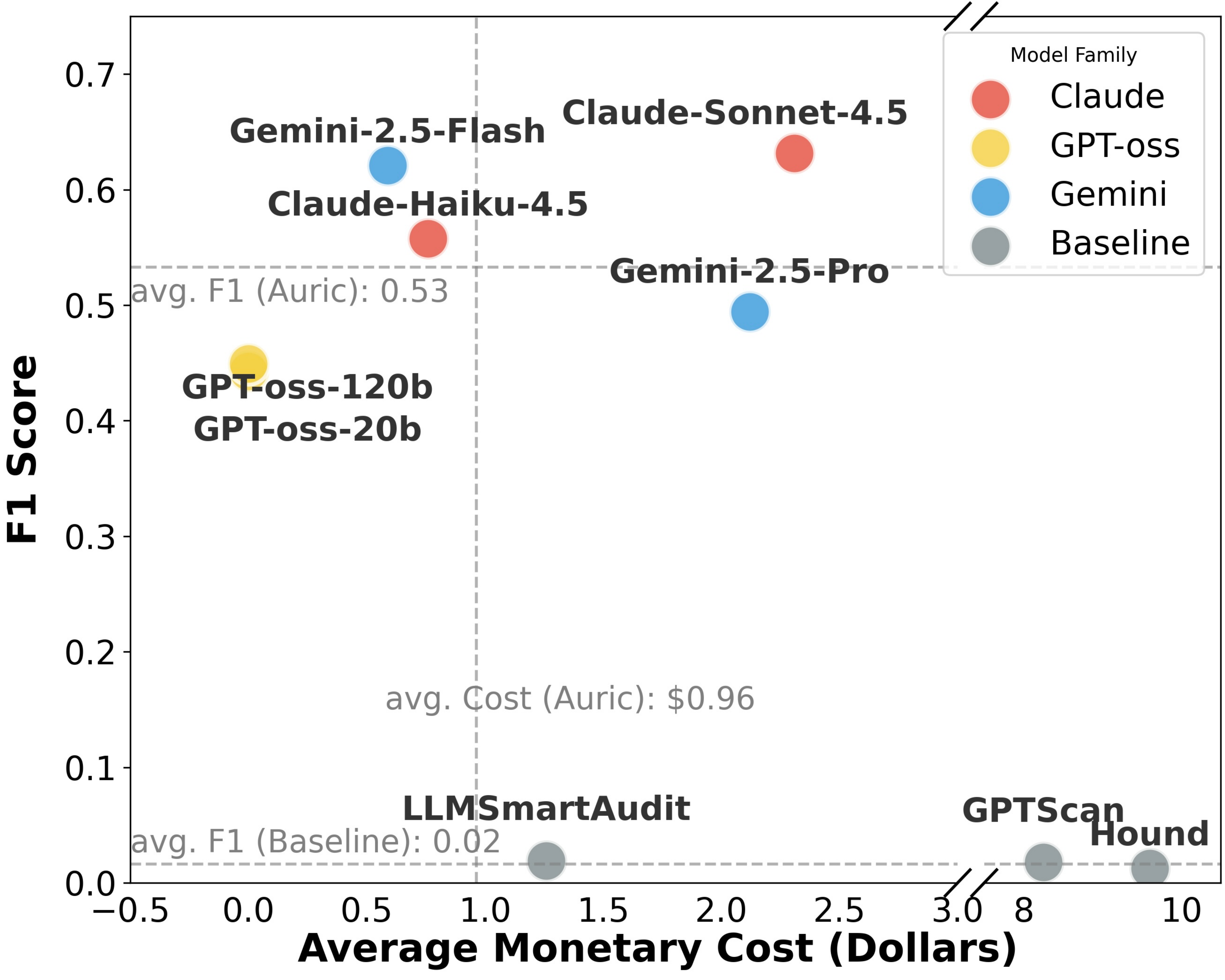}
        \caption{F1 vs. Monetary Cost.}
        \label{fig:money}
    \end{minipage}%
    \hfill
    \begin{minipage}[t]{0.495\linewidth}
        \centering
        \includegraphics[width=\textwidth]{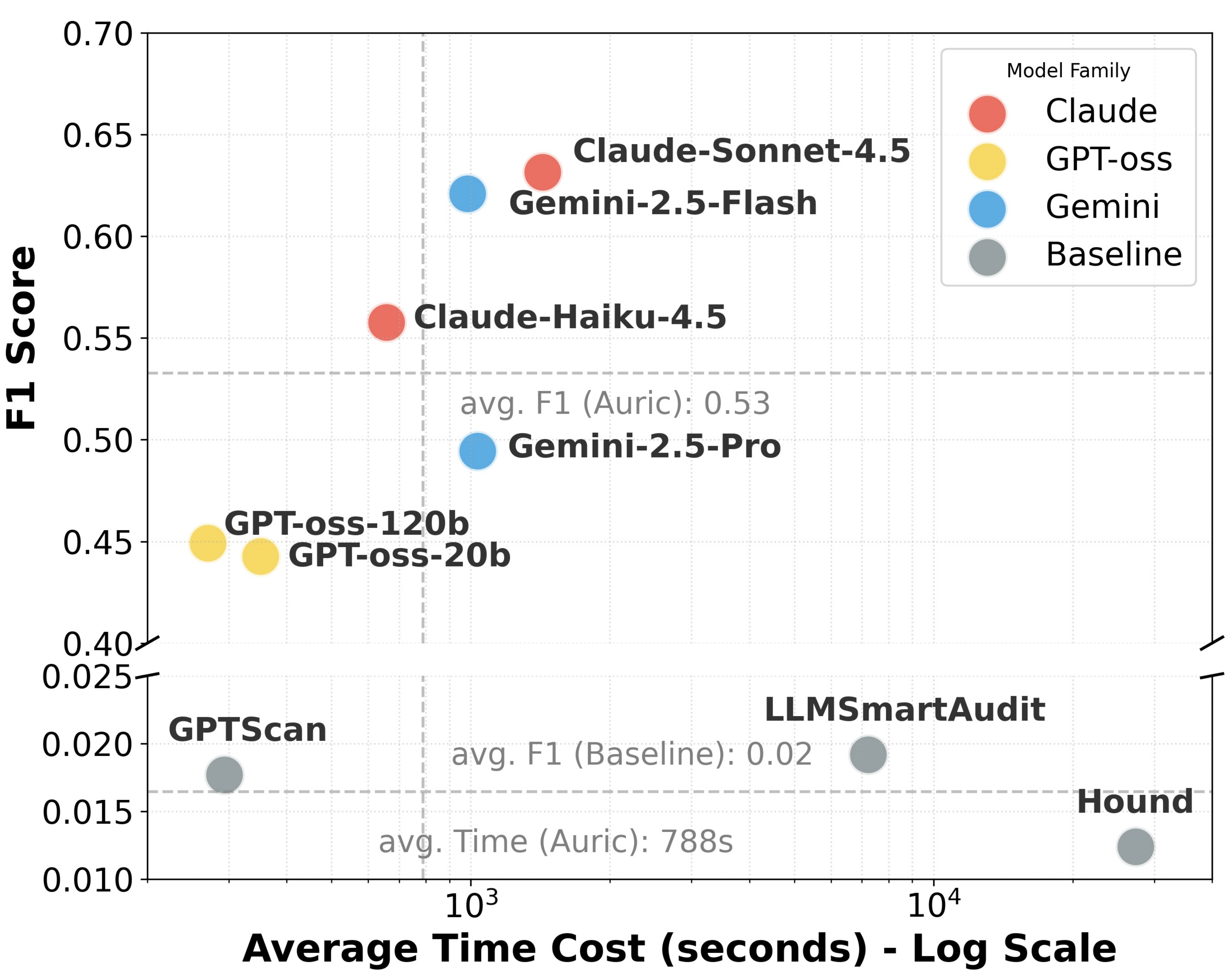}
        \caption{F1 vs. Time Cost.}
        \label{fig:time}
    \end{minipage}
\end{figure}

High false positive rates are the primary barrier to the adoption of automated auditing tools. RQ1.2 assesses whether {\tool} effectively mitigates this issue. The results of baselines and {\tool} are detailed in Table~\ref{tab:baseline-perf} and Table~\ref{tab:rq_d3_performance}, respectively. We also present intuitive comparisons between baselines and {\tool}, focusing on F1 Score and monetary/time cost, in Figure~\ref{fig:money} and Figure~\ref{fig:time}.

Existing academic tools suffer from catastrophic noise levels. As shown in Table~\ref{tab:baseline-perf}, \textit{GPTScan} and \textit{LLMSmartAudit} exhibit the false positive rates of 97.5\% and 99.01\%, respectively. This means for every 100 alerts they generate, only 1--3 are valid, rendering them practically unusable for human auditors.
In stark contrast, as shown in the top half of Table~\ref{tab:rq_d3_performance}, regardless of the LLMs adopted, the false positive rate does not exceed 50\%. Furthermore, in the best-case scenario, the false positive rate can be as low as 31.27\%, which is achieved using GPT-oss-20B as the underlying LLM. This demonstrates the superiority of our proposed method. Interestingly, the two lightest LLMs, \textit{i.e.,} GPT-oss-20B and Gemini-2.5-Flash, have the lowest false positive rates, aligning with an existing comprehensive evaluation~\cite{bi2025gpt}. In contrast, larger models with stronger associative reasoning capabilities may over-interpret benign code patterns as complex vulnerabilities, leading to higher false alarm rates.
Crucially, this proves that {\tool} enables effective auditing via local deployment, effectively resolving data privacy concerns without requiring high-end GPU clusters.

Among baselines, \textit{Hound} shows a significant decrease in false positive rate compared to the other two, thus achieving an order-of-magnitude improvement in precision. However, the statistical results also indicate that \textit{Hound} has an extremely high false negative rate (>99\%), meaning that \textit{Hound} is relatively \textit{conservative}, \textit{i.e.,} it only reports vulnerabilities with high confidence. Therefore, the f1-scores of the three baselines remain around 0.01. In contrast, {\tool}, regardless of the LLMs used, achieved an f1-score above 0.45, with the highest reaching 0.63 (powered by Claude-Sonnet-4.5). This once again demonstrates the balance achieved by our method between detection rate and false positive rate.

\textit{\textbf{Answer to RQ1.} {\tool} demonstrates superior effectiveness by achieving an 86.7\% detection rate for complex business logic vulnerabilities, significantly outperforming state-of-the-art tools like the Claude agent (51.1\%). Furthermore, it effectively mitigates noise by maintaining a low false positive rate between 31.27\% and 49.12\%, a stark contrast to existing academic tools which suffer from rates as high as 99\%.}

\subsection{RQ2: Cost-Efficiency}
\label{subsec: answer rq2 costEfficiency}
In this section, we evaluate the practicality of {\tool} for real-world adoption by quantifying its operational efficiency. Beyond achieving a high detection rate, {\tool} also demonstrates superior efficiency in terms of both monetary and time costs.

Our statistics on $\mathcal{D}_1$ and $\mathcal{D}_2$ indicate that {\tool} incurs a cost of only \$0.22 per project (see Table~\ref{tab:performance on D1, D2}), representing a 68.1\% reduction compared to the second most affordable option, \textit{GPTScan}.
We also observe a favorable cost profile for {\tool} compared to industry standards in $\mathcal{D}_3$. For instance, the static analysis tool \textit{Hound} incurs an average cost of \$9.60 per audit. In comparison, {\tool} offers a range of cost options: the high-precision Claude-Sonnet-4.5 configuration costs \$2.31, while the efficient Gemini-2.5-Flash configuration reduces this only to approximately \$0.59 per audit.

Regarding execution time, {\tool} demonstrates efficiency gains over traditional comprehensive analysis. As Table~\ref{tab:baseline-perf} illustrates, while tools like \textit{Hound} and \textit{LLMSmartAudit} required several hours, around 7.6h and 2h, respectively, to complete the benchmark tasks, {\tool} completed the same audits in under 30 minutes across all configurations (\eg, 4.5 minutes for GPT-oss-120B and 23.8 minutes for Claude-Sonnet-4.5). This efficiency facilitates more frequent security checks during the development lifecycle. As for \textit{GPTScan}, though with a relatively low f1-score, it could complete given auditing tasks in five minutes on average. We note that its low time overhead is attributed to its adoption of GPT-3.5-turbo, OpenAI's fastest model.

\textit{\textbf{Answer to RQ2.} {\tool} achieves exceptional cost-efficiency with an average audit cost of just \$0.22 per project and at most \$2.31 per 10K LOC.
Additionally, it completes scans in an average of 190 seconds on real-world projects. {\tool} operates orders of magnitude faster and cost-efficiently than LLM-based auditors like Hound.}

\subsection{RQ3: Ablation Study}
To evaluate the contribution of Verifier (\S\ref{subsec: implementation:verifier}), we conducted an ablation study by comparing the performance of {\tool} against the version without the Verifier, denoted as ${\tool}_{w/o-V}$. 
Results are shown in Table~\ref{tab:rq_d3_performance}, removing the Verifier results in a consistent increase in the detection rate and a corresponding decrease in the false negative rate across all underlying LLMs. For instance, with Claude-Sonnet-4.5, the detection rate rises from 78.92\% to 92.45\%. This phenomenon suggests that the Auditor is highly effective at vulnerability hypotheses generation and tends to be over-sensitive, flagging a wide range of potential issues to minimize risks of missing critical vulnerabilities.

However, this increase in detection capability comes at a high cost to precision. The data reveals a sharp deterioration in the false positive rate and precision for ${\tool}_{w/o-V}$. In the case of Claude-Haiku-4.5, the false positive rate surges from 49.12\% to 86.16\%, causing precision to plummet from 50.88\% to a mere 13.84\%. This drastic shift indicates that while the Auditor successfully identifies real vulnerabilities, it also introduces a substantial amount of noise. The Verifier is therefore proven to be essential for filtering out these false alarms, acting as a critical quality gate that distinguishes actual logic flaws from benign code patterns.

Ultimately, the trade-off provided by the Verifier is justified by the overall performance metrics. Although ${\tool}_{w/o-V}$ achieves a higher detection rate, its low precision results in significantly lower F1 Scores compared to the complete {\tool} (\textit{e.g.,} 0.3707 vs. 0.6316 for Claude-Sonnet-4.5). The Verifier effectively suppresses the noise, sacrificing a small margin of recall, likely due to rejecting low-severity true positives, to achieve a much higher standard of reliability. Crucially, our analysis of the filtered true positives reveals that this sacrifice is strategically acceptable: among valid vulnerabilities incorrectly filtered out, 91.89\% were \textit{Medium} or \textit{Low} severity, while only 8.11\% were High severity, and importantly, zero Critical vulnerabilities were missed. This confirms that the dual-stage architecture is vital for practical auditing, where a high volume of false positives would render the tool unusable for developers.

\textit{\textbf{Answer to RQ3.} The Verifier is indispensable for filtering the substantial noise generated by the initial detection phase, as evidenced by the fact that removing it causes the false positive rate to surge from 49.12\% to 86.16\% (using Claude-Haiku-4.5). While the Verifier slightly reduces the raw detection rate, it drastically improves overall reliability, raising the f1-score from 0.3707 to 0.6316.}

\subsection{RQ4: Case Study}
To validate the practical utility of {\tool} beyond benchmarking datasets, we apply {\tool} on real-world projects to try to discover zero-day vulnerabilities in production environments.

\subsubsection{Discovery of Zero-Day Vulnerabilities}
\label{subsec: rq4_zero-days}
We deployed {\tool} (supported by Claude-Sonnet-4.5) in two distinct real-world scenarios, demonstrating its versatility and effectiveness in identifying unknown threats.

\noindent\textbf{Industrial Deployment and Collaboration.}
To evaluate practical utility in real-world settings, we deployed {\tool} in two distinct industrial scenarios. 
First, we integrated {\tool} into a prominent DeFi protocol with a Total Value Locked (TVL) exceeding \$400 million. During the pre-deployment audit phase, {\tool} successfully identified \textit{three previously unknown critical vulnerabilities}, all of which could lead to direct financial losses, and they were about cross-contract reentrancy and token integration inconsistency vulnerability. These findings were confirmed by the protocol's core team and subsequently patched before the mainnet upgrade, preventing potential financial losses.
Furthermore, in collaboration with an industry-leading smart contract auditing firm, {\tool} functioned as an auxiliary auditor on a separate client project. It has successfully flagged \textit{four valid medium-severity issues} that were initially overlooked during the preliminary manual review. While human auditors overlooked the liquidation bonus edge case, {\tool} successfully detected the vulnerability by correctly simulating a sandwich attack scenario. This demonstrates {\tool}'s value as a reliable ``second pair of eyes'' to augment human expertise.

\noindent\textbf{Public Auditing Contests.}
We deployed {\tool} in three live Sherlock auditing contests, where it automatically generated reports identifying a total of 131 potential vulnerabilities. So far, official results have been finalized for only one contest. In this contest, we submitted the single issue with the highest confidence, which was confirmed by the judges as \textit{high-severity}. The vulnerability involved malicious input: {\tool} detected insufficient input parameter checks, hypothesizing that an attacker could deploy a malicious contract to bypass privilege checks and execute the exploit. We will provide updates as results from the other contests become available. This contribution ranked {\tool} approximately \textit{19th out of 548} participating human auditors, placing the automated agent within the top 4\% of the leaderboard, a significant milestone for autonomous auditing systems.

\subsubsection{Reproduction of High-Value Exploits}
To further assess {\tool}'s capability in detecting real-world threats, we conducted a retrospective analysis on 20 high-value DeFi exploits that occurred after June 2025. This timeframe ensures that the vulnerabilities were not part of {\tool}'s training data, thereby testing its ability to identify zero-day vulnerabilities. We provided the pre-exploit source code of the victim contracts as input to {\tool} and analyzed the generated audit reports.

Remarkably, in \textit{17 out of the 20 cases}, {\tool} successfully generated reports that accurately pinpointed the root cause functions and described exploit strategies identical to those employed by the actual attackers. For the remaining three cases, our manual inspection revealed that the vulnerabilities relied on off-chain components or external oracle manipulation not visible within the provided scope. 
These results suggest a significant potential for harm reduction: had these DeFi protocols utilized {\tool} prior to deployment, \textit{17 of these 20 attacks} could have been preemptively identified and mitigated. In an optimistic scenario, this would have prevented a cumulative financial loss of approximately \$384 million.

\subsubsection{Case Study: Balancer V2 Exploit}
\label{subsec:rq4 case study}
We detail {\tool}'s performance on the \$128M Balancer V2 exploit~\cite{balancerExploit}, a complex attack driven by a ``Precision Loss Cascade.'' The vulnerability allowed attackers to compound negligible rounding errors via 65+ micro-swaps in a single \texttt{batchSwap} transaction.

{\tool} successfully identified this critical bug through its unique neuro-symbolic pipeline. First, the Profiler grouped the \texttt{Composable-\\StablePool} logic with the \texttt{Vault}'s execution interface. Guided by the ``Integer Division Rounding in Fixed Point Operations'' knowledge pattern, the Auditor flagged the \texttt{\_upscaleArray} function, warning that \texttt{mulDown} operations could generate ``Accumulated Dust.'' Crucially, while LLMs struggled with the arithmetic, the Z3 integration mathematically proved that ``Calculations Near Boundaries'' (specifically 8-9 wei inputs) caused significant invariant divergence when iterated. Finally, the Synthesis step reconstructed the full attack chain. By validating this precision discontinuity via Z3 while retaining semantic guidance, {\tool} detected a vulnerability that evaded both static analyzers and standard LLM auditors.

\textit{\textbf{Answer to RQ4.} {\tool} successfully identified 3 critical zero-days in a \$400M protocoll; assisted auditors with real-world auditing; and ranked in the top 4\% of a Sherlock contest. Retrospective analysis confirms it could have preemptively detected 17 of 20 recent major hacks, demonstrating robust capability in securing real-world DeFi ecosystems.}

\section{Related Work}

\noindent\textbf{Static Analyzers for Smart Contracts.}
Static analysis inspects Solidity or EVM bytecode without executing it to flag known weakness patterns and contract-quality issues. These tools provide fast, repeatable checks and actionable diagnostics, forming the first line of defense in most auditing pipelines. SMARTCAT~\cite{bosi2025following} detects price manipulation attacks via bytecode analysis, and \text{AVVerifier}~\cite{sun2024allyoutokens} identifies vulnerabilities of the address verification process in smart contracts. Key analyzers include Oyente’s~\cite{luu2016making} symbolic execution for reentrancy and TOD patterns, Securify’s~\cite{tsankov2018securify} compliance/violation patterns over bytecode facts, and Slither~\cite{slither} for building intermediate representations and running taint/data-flow detectors over the AST to identify issues. While they define the operational baseline for security analysis, they suffer from imprecision, limited ability to model dynamic behaviors, and significant maintenance and triage overhead in modern DeFi codebases. 

\noindent\textbf{Dynamic Analyzers for Smart Contracts.}
Dynamic analysis executes contracts to generate concrete PoCs, complementing static checks by validating exploitability. Greybox fuzzers such as sFuzz~\cite{nguyen2020sfuzz}, ContractFuzzer~\cite{jiang2018contractfuzzer}, and Smartian~\cite{choi2021smartian} leverage execution feedback to improve path coverage. However, fuzzers cannot guarantee completeness and frequently miss deep, state-dependent logic errors. Furthermore, they require intricate environment modeling and formal specifications, creating operational bottlenecks that hinder scalability in real-world development pipelines.

\noindent\textbf{LLM-based Smart Contract Auditing.}
Recent work increasingly frames auditing as language-guided program analysis, where LLMs coordinate code understanding, hypothesis generation, and evidence gathering. Representative multi-agent systems, including LLM-SmartAudit~\cite{wei2024llmaudit} and Hound~\cite{hound}, coordinate collaborative agents to scan repositories, significantly enhancing coverage. While these approaches improve coverage and analyst throughput, they remain sensitive to LLM planning reliability, and they can be slow and costly due to extensive tool calls and multi-turn deliberation. AuditGPT~\cite{xia2024auditgpt} decomposes ERC~\cite{ercstandards} compliance into rule-focused checks, enabling systematic, explainable assessments with reduced hallucination risk. However, it may fail to generalize to the bespoke, cross-contract business logic found in complex DeFi protocols, necessitating more robust neuro-symbolic architectures.

\section{Discussion}
The ultimate vision for {\tool} is to seamlessly embed expert-level security into the modern DeFi development lifecycle, effectively dismantling the barriers of prohibitive costs, alert fatigue, and data privacy risks. By achieving comprehensive audits for as low as \$0.59 without compromising detection quality (F1=0.62), {\tool} shifts the paradigm from episodic ``audit-once'' checks to continuous DevSecOps. Crucially, this high-frequency auditing is sustained by our aggressive false positive filtration (\S~\ref{subsec: implementation:verifier}), which prioritizes precision to ensure developers treat human attention as a scarce resource rather than filtering through noise. 

Beyond operational efficiency, {\tool} addresses critical institutional needs for data sovereignty. Unlike cloud-based solutions that expose sensitive code to external servers, {\tool}'s model-agnostic architecture supports local deployment on consumer-grade hardware (\eg, \textit{GPT-oss-120B}). This capability offers a \textbf{privacy-first solution}, enabling rigorous, air-gapped security analysis that safeguards proprietary logic within a secure perimeter.

\section{Limitations and Future Work}
Despite the promising results, {\tool} has limitations that guide our future research directions:

\noindent\textbf{Language Support and Static Analysis Dependency.}
Currently, {\tool}'s contextual profiling (\S~\ref{subsec: M1 methodology-profiler}) relies on Slither~\cite{slither} for dependency graph generation, restricting our scope to Solidity smart contracts. This is a limitation of the underlying static analysis infrastructure rather than the agentic framework itself. In future work, we plan to modularize the graph generation component, integrating language-agnostic tools like tree-sitter~\cite{tree-sitter} or developing custom parsers to support other contract languages such as Vyper, Rust (Solana), and Move (Aptos/Sui).

\noindent\textbf{Knowledge Base Maintenance.}
The effectiveness of our \textit{Remind} mechanism relies on the currency of the Knowledge Base ($\mathcal{K}$). While the system performs well on known vulnerability classes, it may miss entirely novel attack vectors (zero-day classes) absent from the database. However, the maintenance overhead is minimal. Since $\mathcal{K}$ is structured as a file system rather than a fine-tuning dataset, updating the system requires only the ingestion of new vulnerability reports as text documents, which we commit to maintaining as an open-source resource.

\noindent\textbf{Scope of Simulation.}
Our current adversarial simulation focuses on atomic transaction sequences. Complex governance attacks that unfold over weeks (\eg, proposal manipulation) or off-chain phishing vectors are currently outside the system's modeling scope. Future versions will explore integrating event-driven simulation environments to model long-term protocol state changes.

\section{Conclusion}
This paper introduces {\tool}, a smart contract auditing agent designed to resolve the scalability-precision trade-off in DeFi security. Unlike unscalable manual auditing or hallucination-prone LLM tools, {\tool} employs a ``Plan-Remind-Solve'' workflow. Moreover, it utilizes the Contextual Profiling to manage context limits and a False Positive Filtration layer to rigorously verify findings.

Evaluations demonstrate {\tool}’s superiority over state-of-the-art baselines. Powered by Claude-Sonnet-4.5, it achieves an 86.7\% detection rate on high-value exploits—substantially outperforming the Claude baseline (51.1\%), while drastically reducing false positives. The system is highly cost-efficient, operating at \$0.22 per project with lightweight models or \$2.31 per 10K LOC for maximum precision. Practically, {\tool} reproduced 17 of 20 post June 2025 exploits and identified confirmed zero-day vulnerabilities in live protocols with over \$400 million TVL. {\tool} thus provides a production-ready solution for integrating continuous, expert-level security into the DeFi development lifecycle.

\newpage
\appendix
\section{Ethical Considerations}
\label{appendix:ethical consideration}

The primary goal of {\tool} is to secure smart contracts before deployment, preventing financial losses in the DeFi ecosystem. However, we recognize that automated auditing tools pose a dual-use risk that attackers could potentially misuse them to find zero-day vulnerabilities in live protocols.
To prevent such abuse, we implement a controlled distribution strategy. Instead of open-sourcing the core exploit generation engine, we offer {\tool} exclusively as an API service. Access is restricted to verified developers and auditing firms subject to a vetting process. Additionally, we strictly adhere to responsible disclosure protocols. Any new vulnerabilities discovered by {\tool} are privately reported to the affected project teams, ensuring they are remediated before any public information is released.

Regarding the datasets employed in our evaluation, we explicitly clarify that all analyzed smart contracts and historical exploit reports (referenced as $\mathcal{D}_1$, $\mathcal{D}_2$, and $\mathcal{D}_3$) are derived exclusively from publicly available sources, including the Ethereum mainnet and open auditing contest platforms (\eg, Sherlock). These projects are open-source and globally accessible to the general public; therefore, our data collection and analysis processes involve no unauthorized usage of private intellectual property or non-public codebases.

\section{Generative AI Usage}
During the preparation of this work, the authors utilized LLMs, specifically Gemini-3-Pro, to assist in refining the clarity of the text and drafting specific sections, including the Abstract, Discussion, and Conclusion. These tools were also employed to generate LaTeX formatting templates.
The authors maintained full responsibility for the accuracy and originality of the content. All AI-generated text was critically reviewed, verified, and edited by the authors to ensure it accurately reflects the research findings and experimental data. No AI tools were used to manufacture, falsify, or scientifically alter the experimental results or data presented in this paper.

\section{Prompt}

This Appendix provides several parts of our prompts.
\subsection{Prompt 1}
\label{appendix:prompt1}

\# Instructions

Please analyze the provided code using the following Chain-of-Thought process:

1.  **Semantic Analysis**: Briefly describe the role of each contract in this batch (e.g., Vault, Oracle, Strategy).

2.  **Sensitive Operation Scan**: Identify functions that perform critical operations, specifically:
    *   Asset transfers (ETH/ERC20).
    *   State changes to critical variables (e.g., balances, exchange rates).
    *   Low-level calls `call`, `delegatecall`) or external interactions.
    *   Complex arithmetic logic (e.g., yield calculation).
    
3.  **Heuristic Filtering**: Filter out standard/safe boilerplate (e.g., standard ERC20 `approve` logic) unless it deviates from the norm. Focus on custom business logic.

4.  **Selection**: Select functions that might be vulnerable to attacks like Reentrancy, Access Control bypass, or Logic Errors.

\subsection{Prompt 2}
\label{appendix:prompt2}

\# exploration

Actively search for unfamiliar patterns, recent exploits, and protocol-specific risks;

Use knowledge base as starting point, NOT boundary;

Generate attack hypotheses beyond documented vulnerabilities;

Model economic incentives: when is attacking profitable?

\noindent
\# Risk Coverage

Systematically examine:

Logic: edge cases (zero, max, empty), state transitions

Roles: admin, user, attacker, colluding parties

States: paused, upgrading, emergency mode

Interactions: reentrancy, callbacks, delegatecall chains

Temporal: timestamp manipulation, front-running, MEV

Economic: arbitrage, liquidation cascades, governance attacks

\subsection{Prompt 3}
\label{appendix:prompt3}

\# Instructions

Sort the tasks using the following Chain-of-Thought:

Consequence Analysis: For each task, evaluate the Potential Serious Consequence if the function contains a vulnerability.

Critical: Direct loss of all TVL, permanent freezing of funds.

High: Theft of user yield, manipulation of governance parameters.

Medium/Low: Griefing, temporary denial of service.

Weighted Prioritization: Combine the provided Contract Importance Score (structural centrality) with your Consequence Analysis.

Rule: A critical financial function in a highly-scored contract is Priority \#1.

Rule: A critical function in a low-score contract usually outranks a trivial function in a high-score contract.

Ordering: Rank the Task IDs from highest priority to lowest.

\newpage
\bibliographystyle{ACM-Reference-Format}
\bibliography{reference}

\end{document}